**Title**

Integrating spatially-resolved transcriptomics data across tissues and individuals: challenges and opportunities

**Authors**

Boyi Guo[1,*], Wodan Ling[2,*], Sang Ho Kwon[3,4,5], Pratibha Panwar[6,7,8], Shila Ghazanfar[6,7,8,+],
Keri Martinowich[3,4,9,10,11,+], Stephanie C. Hicks[1,12,13,14,+]

**Affiliations**

1. Department of Biostatistics, Johns Hopkins Bloomberg School of Public Health, Baltimore, MD, USA
2. Division of Biostatistics, Department of Population Health Sciences, Weill Cornell Medicine, NY, USA
3. Lieber Institute for Brain Development, Johns Hopkins Medical Campus, Baltimore, MD, USA
4. Solomon H. Snyder Department of Neuroscience, Johns Hopkins School of Medicine, Baltimore, MD, USA
5. Biochemistry, Cellular, and Molecular Biology Graduate Program, Johns Hopkins School of Medicine, Baltimore, MD, USA
6. School of Mathematics and Statistics, The University of Sydney, NSW 2006, Australia
7. Sydney Precision Data Science Centre, University of Sydney, NSW 2006, Australia
8. Charles Perkins Centre, The University of Sydney, NSW 2006, Australia
9. Department of Psychiatry and Behavioral Sciences, Johns Hopkins School of Medicine, Baltimore, MD, USA
10. Johns Hopkins Kavli Neuroscience Discovery Institute, Baltimore, MD, USA
11. Department of Biomedical Engineering, Johns Hopkins University, Baltimore, MD, USA
12. Center for Computational Biology, Johns Hopkins University, Baltimore, MD, USA
13. Malone Center for Engineering in Healthcare, Johns Hopkins University, Baltimore, MD, USA

*co-first authors
+co-corresponding authors

**Correspondence**

Shila Ghazanfar
shila.ghazanfar@sydney.edu.au

Keri Martinowich
keri.martinowich@libd.org

Stephanie C. Hicks
shicks19@jhu.edu

# Abstract

Advances in spatially-resolved transcriptomics (SRT) technologies have propelled the development of new computational analysis methods to unlock biological insights. As the cost of generating these data decreases, these technologies provide an exciting opportunity to create large-scale atlases that integrate SRT data across multiple tissues, individuals, species, or phenotypes to perform population-level analyses. Here, we describe unique challenges of varying spatial resolutions in SRT data, as well as highlight the opportunities for standardized preprocessing methods along with computational algorithms amenable to atlas-scale datasets leading to improved sensitivity and reproducibility in the future.

# Main

Comprehensive atlases of molecular profiles with spatial resolution have the power to provide new insights into human health and disease, which can transform the future of medicine via improved diagnostics and targeted therapies [1,2]. Recent commercialization has led to broad accessibility and hence collection of substantial amounts of spatially-resolved transcriptomics (SRT) data, signifying a new era for spatial cellular atlases and charting the unknown territory of life science [3]. These technologies enable mapping of heterogeneous cell populations *in situ* to tissue architectures, equipping investigators to study the relationships between structure and biological activities [4]. Computational tools and analytic strategies that can fully exploit the atlas-scale SRT data and increase the power to detect small biological signals are critically needed [5–7]. However, integrating multiple tissues [8], developmental stages [9,10], species [11,12], or phenotypes [13] to perform population-level analyses faces new and unique challenges.

In contrast to single-sample analyses [14], performing population-level analyses with an integrated set of SRT samples has the potential to identify spatially-dependent commonalities and differences at the population-level across disease states or conditions such as Alzheimer's disease [15,16], schizophrenia [17], and cancer [18,19]. Here, we discuss the computational challenges involved in analyzing an integrative spatial atlas across tissues and individuals with a focus on the existing computational strategies currently available as well as future opportunities for development. We focus on the challenge of integrating SRT samples where observations are measured at different levels of spatial resolution due to inherent capabilities and limitations of the employed technologies. We illustrate that varying levels of resolution combined with differences in the features captured can lead to spurious findings in downstream analyses, such as dimensionality reduction. These problems are exacerbated by challenges faced in bulk and single-cell/nucleus RNA-sequencing (sc/snRNA-seq) data, such as sparsity and noise [20]. Finally, we summarize the state-of-the-art methods for integrating multiple SRT samples to perform population-level analyses.

## From bulk to single-cell and spatial resolution

As the implementation of integration has become commonplace as the number of SRT datasets has increased, the value of reliably identifying shared or distinctive cellular features across these data sets has been demonstrated. Unwanted variation across samples or datasets, which are ubiquitous across most sequencing modalities ranging from bulk to single-cell data [21,22], and are routinely referred to as *batch effects*, are an inevitable challenge faced during integration [20,23]. This undesired heterogeneity usually comes from artifacts such as differences in handling protocols, library preparation and sequencing platforms. Therefore, correcting for batch effects is a major goal of integration. Examples for how to correct for batch effects in bulk RNA-seq include the use of statistical modeling to adjust for sample-level differences [24–26] along with the use of control genes [27].

In contrast to bulk RNA-seq, which measures gene expression in one sample that is averaged across measured cells, scRNA-seq measures gene expression across thousands to millions of cells and introduces more heterogeneity in the gene expression space. Therefore, as we moved from bulk to single-cell resolution, one type of integration strategy that was developed for scRNA-seq experiments was to identify groups of cells that share similar expression patterns across batches (called *anchors*). Broadly these approaches use similarity-based methods in a reduced dimension space, such as mutual nearest neighbors (MNN) [28], Harmony [29], and canonical correlation [30]. The key idea is that similar cells should follow a common distribution in the latent space regardless of batch. As an extension of dimension reduction methods, generative models effectively help capture nonlinear characteristics of batch effects and systematic biological signals, such as improving the exhaustiveness of artifact elimination [31].

However, a prominent feature of scRNA-seq data is that the measured observation, namely gene expression in one cell, is the same, in principle, across all observations measured in multiple scRNA-seq experiments. With SRT, the observations that we measure within a tissue may be the same, but the resolution of observations across multiple samples may not be the same (**Figure 1**). Therefore, while these integrative methods developed for bulk and scRNA-seq experiments demonstrate significant success when integrating bulk and single-cell data, it remains unclear how well these methods will work for SRT data due to intrinsic differences in experimental protocols and the biological context of generated data. For example, this motivates the use of alternative pieces of information, such as anatomical landmarks [32,33], to assist in the construction of population-level spatial atlases, but these are not always relevant, for example with cancer tissue.

## Inconsistent spatial and biological resolutions challenge cross-technology integration

'Spatially-resolved transcriptomics' [34], is used as an umbrella term for multiple distinct technologies that can measure spatial gene expression [35]. However, due to intrinsic differences in how they measure spatial gene expression, these data have unique computational and biological properties that make using integration strategies developed for scRNA-seq data analysis challenging.

For example, the units in which we measure observations, namely individual cells or groups of cells, referred to as *observational units*, vary substantially across the SRT platforms. In image-based, targeted, *in situ* transcriptomic profiling, such as MERFISH [36] or Xenium [37], gene expression is captured from a targeted subset of genes at molecule-level resolution, where the molecules are aggregated together to computationally infer the "cellular" observational unit using cell segmentation algorithms. In contrast, non-targeted RNA capture and sequencing profiling, such as Slide-seq [38] or Visium [39], captures RNA on an array-based platform at different resolutions, including "near-cellular" (such as 55 µm spots on the Visium platform) or "sub-cellular" (such as 2 µm grids on the VisiumHD platform [40]). Integration of data generated across technologies with different observational units requires special attention.

Unlike the concept of the observational unit whose distinction across SRT technologies is well acknowledged [3], the heterogeneity in biological content being profiled across observations generated from the same SRT technology is often overlooked. Sc/snRNA-seq protocols often employ cell dissociation techniques to isolate individual cells or nuclei. When processed correctly, data observations have a uniform and biologically meaningful unit (referred to as *biological unit* hereafter), cell or nuclei, across samples and studies. However, because the profiling happens *in situ*, this property is often missing in SRT data. For example, with sequencing-based SRT technologies, the profiling within the observational unit is constrained by physical size, e.g. spots or grids, so the generated data observation frequently does not maintain a uniform biological content, and hence the *biological unit* of data observations varies widely. Specifically, the cellular structures being captured across observations could include both soma of cells and the extracellular space between multiple cell bodies (**Figure 1**). Inconsistency in biological units in SRT datasets greatly challenges the fundamental assumption that many integration methods for sc/snRNA seq data depend on, namely that each observation is an individual cell. This can lead to spurious results or biases in fundamental data preprocessing steps, such as data normalization [41], quality control [42], and in turn propagate through downstream integration steps.

Even the single-cell resolution image-based SRT technologies may suffer from the inconsistency of biological units across data observations. Despite individual SRT tissue sections being conceptually treated as 2D objects, each tissue section has a 3D structure, meaning that the tissue section has some dimension into the Z-plane. Depending on their orientation, it is possible for cells to be bisected during tissue sectioning. In this scenario a cell will not maintain, full integrity since only a portion of the cell structure is captured (**Figure 1**).

Moreover, many image-based SRT technologies require iterative imaging of small regions of a tissue section. This iterative imaging procedure creates cells that are located at the boundary of images and hence only partially profiled, resulting in variation in captured genes [41,43]. Although the degree of variation in biological units is smaller than sequencing-based SRT data, further research is necessary to understand the downstream impact of these confounders in data analysis and integration.

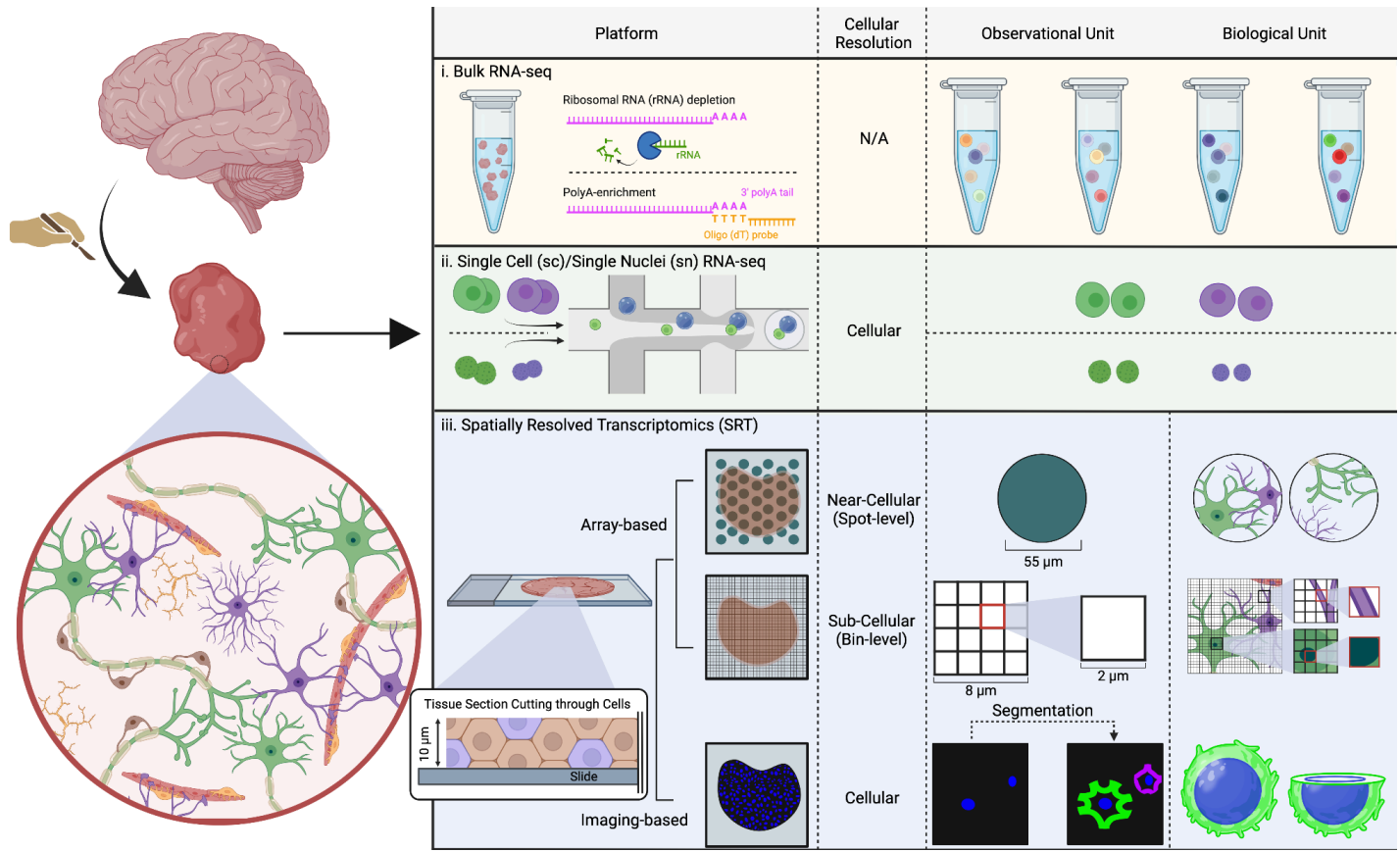

**Figure 1. Schematic of experimental platforms and cellular resolutions across transcriptomics technologies.** Considering three different experimental platforms (i) bulk RNA-sequencing, (ii) single-cell/nucleus RNA-sequencing, and (iii) spatially-resolved transcriptomics, each of these can profile gene expression at different cellular resolutions, including cellular, near-cellular, and sub-cellular. Differences in experimental platforms also have differences in the units being measured, including observational units and biological units, where observational units describe the observations that we measure and the biological units describe the cellular structure that the observation unit captures.

In addition to addressing the inconsistency in observational and biological units when integrating across SRT datasets, another significant challenge is to mitigate the inclusion of divergent sets of assayed genes across platforms or studies. While targeted profiling technologies provide better spatial resolution, they are often limited by the amount of genes profiled. Specifically, targeted panels focus on measuring pre-selected sets of genes that are often tissue- or disease-specific, occasionally with some additional genes that are customized to individual studies. Unlike transcriptome-wide sequencing technologies that capture all genes and hence have a consistent set of genes across studies, the divergent sets of assayed genes from targeted-panels lead to missing gene features when integrating data collected from different studies. Also, the mismatching of gene profiles due to the frequent missing genes issue of the targeted technologies prevents the direct adoption of scRNA-seq methods to integrate data generated from targeted and non-targeted platforms.

# Case study: cross-platform integration using cell type-based anchors

In the following section, we highlight a few examples of how the unique properties of this data generate computational challenges for integrating multiple SRT samples.

Normalization is a critical step in processing transcriptomics data to remove variation due to technical noise. Current normalization practices for SRT data, regardless of platform, are directly adopted from scRNA-seq pipeline. However, whether this practice is uniformly appropriate for the diverse types of SRT data remains unclear. A common practice is to normalize the expression of each gene according to the total number of transcripts detected, often referred to as library size normalization. The library size normalization is based on the assumption that the variation in library size across samples is due to technical reasons. However, due to the inconsistent biological unit across samples, library size could reflect variation attributed to the differences in underlying biology. Hence, library size normalization can overcorrect the technical variation and potentially reduce the biological signal. As a result, downstream tasks, such as spatial clustering to establish functional regions, are significantly impacted [44]. Moreover, Atta et al recently demonstrated that applying library size normalization to targeted SRT data could result in false positive and false negative findings in differential expression testing and spatially variable gene detection [41]. Relevantly, many QC methods rely on descriptive metrics such as library size, total gene detected, which is not robust to the inconsistent biological unit unique to SRT data. Totty et al [42] recently showed that scRNA-seq inspired quality control methods could result in differential removal of data observations across multiple biological functional regions in an undesirable way.

Additionally, cell types, often used as anchors to harmonize multiple datasets in sc/snRNA-seq, could be substantially challenging to be properly defined from both the computational perspective and philosophical perspective. While the main intuition for cell type annotation is that the difference in gene signature between data observations, i.e. cells, is driven by the difference of the cell types, the implicit assumption here is that the data observations are single cells. Nevertheless, for near-cellular and sub-celluar resolution SRT data, such assumption is often violated. Mapping observations with different biological units to the common latent space can create dubious clusters that lack biological meanings and confound cell type-driven anchors for cross-study integration. For example, in near cellular resolution platforms, each observation could contain a homogeneous or heterogeneous cell population, resulting in distinction in biological units across observations beyond simply capturing different numbers of cells. This creates challenges to define cell type-driven anchors and leads to extra cell type clusters, which, in fact, should be merged with existing clearly-defined cell types (**Figure 2B**). In another case, targeted platforms can miss important marker genes. Thus, when integrated with data generated from transcriptome-wide platforms that have a full spectrum of genes, the anchors cannot be accurately established such that some cell types cannot be successfully differentiated (**Figure 2C**). Analytically, the SRT technologies provide an unprecedented opportunity to study the molecular mechanisms underpinning the heterogeneities in functions across tissue regions. Many research questions of interest, investigated using SRT platforms, focus on heterogeneity in gene expression associated with functional regions instead of cell types, requiring a switch of thinking from the cell-type centric to the tissue-centric [45]. As a result, the integration tools and strategies that account for both gene expression space and physical space are highly motivated to establish a common coordinate framework [33].

# State-of-the-art methods

Broadly, methods developed for bulk or scRNA-seq are being widely applied to spatial data, despite the problems outlined above. However, new methods to integrate multiple samples for spatial transcriptomics data have recently been developed. In this section, we outline the modern methods specifically designed for spatial data and give recommendations to data analysts and users of these methods.

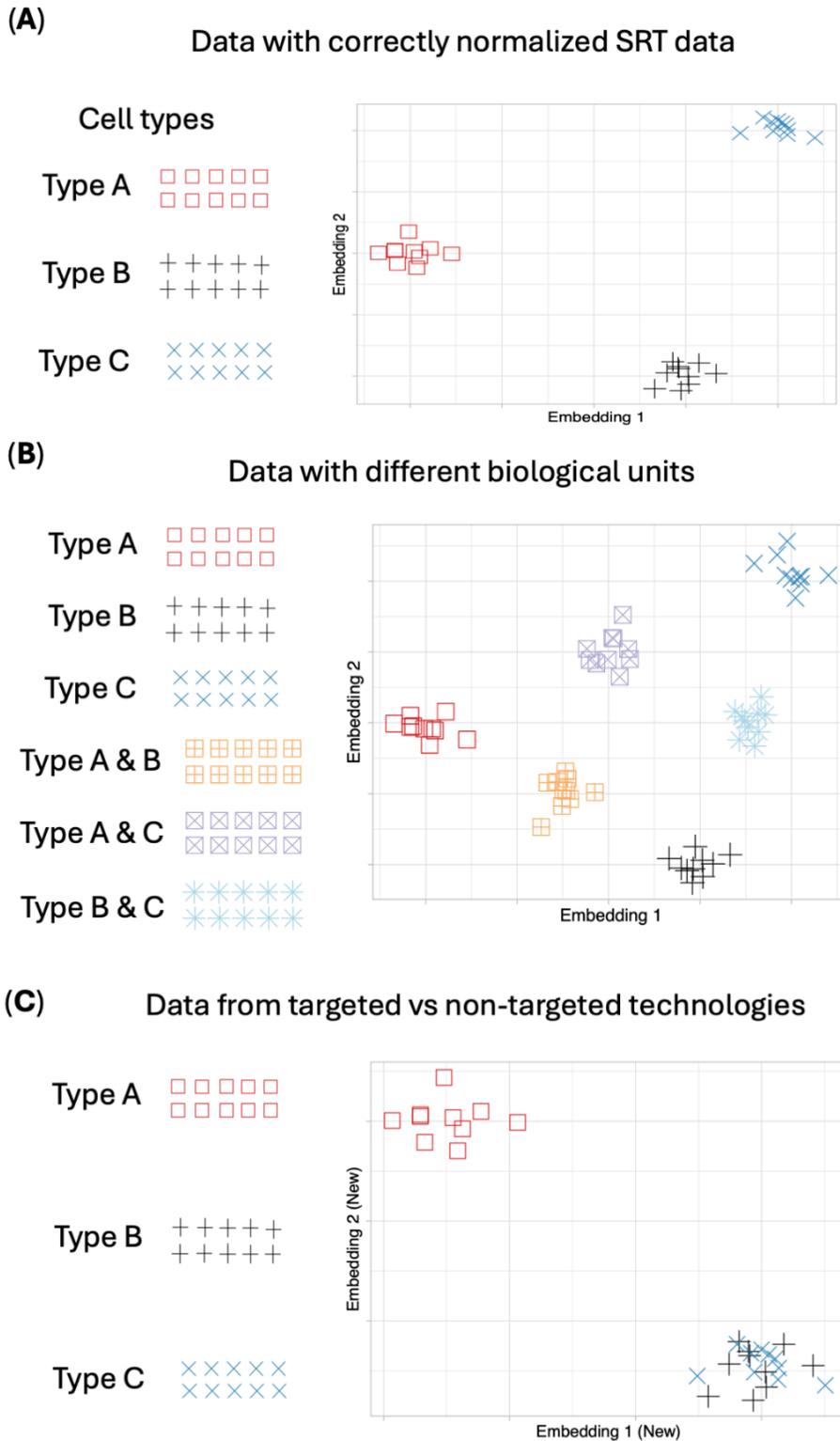

**Figure 2. Schematic of cell-type driven integration in gene expression space across multiple SRT technologies.**
(**A**) With accurate normalization removing technical variation, integration of image-based SRT follows single-cell practice.
(**B**) Mapping observations that have different biological units to a common latent space results in dubious clusters that lack biological meanings and confound cell-type driven anchors for cross-study integration. (**C**) Integrating SRT datasets generated with different gene panels (targeted vs non-targeted) creates challenges to computationally define gene expression space where cell-type based anchors cannot be clearly defined due to missing marker genes in targeted platforms.

**Integration in a physical space**
The first category that we consider is to integrate multiple samples in a physical space. Within this category, we further distinguish approaches based on the type of data being integrated including (i) the alignment of two tissue slices from the same tissue block or from different tissue blocks, but both profiling the transcriptome in a 2D space and (ii) the registration of a set of dissociated single cells to one tissue slice profiling the transcriptome in a 2D space.

Early work of *spatial alignment* were computer-assisted, requiring human input, such as manually defined anatomical landmarks, and computationally relies on the affine transformation, e.g. using iterative closest point algorithm [46], of high-resolution images of samples, e.g. hematoxylin and eosin (H&E) or immunofluorescent images, to address rotations and shifting. Then, various methods were developed to address possible nonlinear distortion, leveraging thin plate spline [47], Gaussian process [48], diffeomorphic metric mapping [49]. Because spatial alignment of tissue images normally requires different degrees of involvement in manual labor, a significant challenge is how to scale it to atlas-scale data sets that contain hundreds of samples. Considerable approaches have been proposed to address this challenge, most involving modeling the entire gene expression profiles accounting for the glocal structure of the spatial unit arrangement, including the two-layer Gaussian process model [48], diffeomorphic metric mapping [49], optimal transport [50], and a graph adversarial matching strategy [51]. These methods seem to be well motivated for the alignment of (i) samples with partial matching, also referred to as spatially heterogeneous samples, (ii) spatial alignment to a reference or template, such as a reference include a predefined Brain atlas [52,53], or (iii) samples across different SRT technologies or possible of various phenotype readouts, such as gene expression and protein expression.

In addition, the *spatial registration* of single cells to a 2D tissue section provides another venue for spatial integration that mitigates analytic challenges due to morphological variation. Specifically, isolated single cells, or possibly spots, are computationally mapped to a spatial common coordinate system/spatial template based on their molecular signature. By mapping all cells isolated from a tissue (either experimentally through scRNA-seq or computationally with SRT data) to a reference or template tissue with uniform morphology, the tissue slices are hence aligned in the physical space with conformable shapes. Developed for the spatial reference assayed with low-throughput technologies, early methods use a set of pre-specified marker genes to anchor cells to a limited number of spatial positions, such as the tessellation of a 2D surface, in Gaussian mixture models [54] or Monte Carlo simulations [55]. To allow non-informative mapping without external reference and improve the spatial resolution of the mapping, advanced machine learning frameworks are adapted, including multiple variations of optimal transport algorithms [56,57], convex optimization using the Jonker–Volgenant shortest augmenting path algorithm [58], and deep neural networks [59,60]. Some spatially-aware deconvolution methods can be also used for the spatial registration of cells [61].

While these methods are promising, it is important to be cautious in the interpretation because it is possible that spurious gene expression correlations are found due to known problems such as 'double dipping' [62]. Given the fast-paced nature of this research area, multi-modal integration approaches might also be useful to avoid the circular problem of double dipping and directly use different data modalities.

**Integration in a latent space**
The second approach ignores (for the most part) the physical space, and focuses on integrating the samples in the latent space. Recent examples are the use of deep learning models [63–65], which combine spatial neighbor networks and graph auto-encoders to learn latent embeddings. In contrast to integration in the physical space, integration in the latent space is anchored in the feature (gene expression) space, while also borrowing information from nearby, physically adjacent cells/spots. Feature space integration has a long history in scRNA data analysis, where gene expression data are projected into the same latent space accounting for batch effects and the downstream investigation is completed in the shared latent space. However, integration that ignores the physical relationships between cells is vulnerable to noise in the data, particularly when the biological signal is small. The spatial information is introduced to the latent space integration to remove the noise, such as in the form of dimension reduction, or possibly in the form of clustering. Distinctly, PRECAST [66] models any arbitrary tissue architecture across multiple tissue samples by factorizing the input into a latent factor with a shared distribution for cell/domain and then using an intrinsic conditional autoregressive component to capture the spatial dependence for spatial clustering. In addition, the spatial information can be also used to smooth out any possible noise. For example, BayesSpace [67] uses majority voting to accomplish spatial smoothing of cluster membership. Once these spatial domains are identified, which normally aligns with anatomical definitions, cells/spots are pseudo-bulked across individual samples, followed by nominal analysis, such as differential expression analysis. However, this pseudo-bulking analysis normally lacks sensitivity for local spatial signals and hence would not be helpful for any granular analysis to study micro-environment.

**Integration in using pseudobulking approaches**
Given the flourishing development of pseudobulking approaches in integration of scRNAseq data [68,69], it is natural to extend this for SRT data where gene expression is aggregated across spots within a spatial domain and a tissue section. As the analysis is conducted at the observational unit of a spatial domain and tissue section, there is no need to address morphological variation. Furthermore, this approach enables existing methods, designed for bulk RNA-sequencing to be used in this setting. For example, pseudobulking SRT data can be used to identify differentially expressed genes (DEGs) across spatial domains with multiple tissue blocks or individuals, which has been successfully applied in human brain tissue including dorsolateral prefrontal cortex (DLPFC) [70,71], locus coeruleus (LC) [72], and the hippocampus (HPC) [73]. In addition, pseudobulking provides a scalable solution to analyzing atlas-scale SRT data, motivated by the underlying "divide-and-conquer" or "map-and-reduce" philosophy.

However, this approach is not necessarily appropriate for all research questions and methods for integrative analysis using SRT data. While this approach sidesteps the challenges due to morphological variation, it also ignores variation due to gene expression patterns varying within a spatial domain as that information is aggregated together into one summary statistic. This loss of information might be particularly important for some downstream analyses, such as identifying spatially variable genes, cellular deconvolution, and cell-cell communication. Pseudobulking or other approaches that summarize features at the sample-level [74] might not be appropriate in these cases.

Furthermore, there are open opportunities to improve the sensitivity and robustness of the statistical models used to perform integrative analyses using pseudobulk data. Current approaches use linear models with empirical Bayes techniques to identify DEGs where the spatial domains are assumed to be discrete. However, more modern models could be considered where the spatial domains are more continuous across a 2D space. In addition, this approach requires labor-intensive human intervention, for example, the need to harmonize the labels corresponding to the spatial domains from unsupervised clustering algorithms. Specifically, the spatial alignment can be addressed via a manual operation, but this can be time-consuming and is prone to human error, leading to potentially unreliable and unreplicable results.

# Discussion

The aim of this work is to both describe the historical context and summarize state-of-the-art strategies to perform population-level analyses that can overcome potential systematic biases in SRT data. The three approaches can broadly be summarized as (i) integration in a physical space, (ii) integration in a latent space, and (iii) integration using pseudobulking or similar approaches. Despite this, approaches developed for scRNA-seq remain widely used in practice. While it remains unclear which type of approach is best to integrate SRT data, there are many ongoing and active efforts to begin comparing these approaches through robust benchmark evaluations.

Furthermore, while much progress has been made towards early strategies, there remain important open challenges to be addressed. For example, when using approaches to integrate tissue sections in a physical space, it remains unclear how to identify the partial overlap and quantify how much area is needed for successful alignment. In addition, smoothing of gene expression is often used implicitly or explicitly as a step in the alignment process before passing to downstream analysis. Further validation to avoid potential over smoothing that eliminates nuanced biological signals in micro environments would be greatly encouraged to avoid introducing computational artifacts. These challenges also relate to potential differences in the amount of biological tissue measured, where one can imagine new spatial platforms capturing a larger tissue area compared to older spatial platforms.

There also remain challenges with the current state-of-the-art strategies, such as the accuracy of cell segmentation, which remains one of the largest challenges with SRT data. Also, while pseudobulking enables the integration across datasets with different observational units, this approach also potentially masks important spatial variation within a given spatial domain. Therefore, we imagine new computational tools being developed that can integrate multiple samples measured with different observational units to take advantage of the full rich information provided by multi-sample SRT datasets.

# Back matter

**Acknowledgements**:
We would like to thank Kasper Hansen and members of the Hicks Lab for their feedback on this commentary. We would also like to thank our collaborators at the Lieber Institute for Brian Development for input and feedback.

**Data availability**:
We used previously published data, which we referenced in the body of the text.

**Abbreviations**:
- sc/snRNA-seq (single-cell/nucleus RNA-sequencing)
- SRT (spatial transcriptomics)
- H&E (hematoxylin and eosin)
- DLPFC (dorsolateral prefrontal cortex)
- HPC (hippocampus)
- LC (locus coeruleus)
- DEGs (differential expressed genes)

**Funding**:
This project was supported by the National Institute of Mental Health [R01MH126393 to B.G., S.H.K., K.M., S.C.H.], and the Chan Zuckerberg Initiative DAF, an advised fund of Silicon Valley Community Foundation [DAF2023-323340 to S.C.H., P.P., S.G.], the National Institute of General Medical Sciences [R01GM151301 to W.L.], and Australian Research Council Discovery Early Career Researcher Awards (DE220100964) funded by the Australian Government to S.G.; Chan Zuckerberg Initiative Single Cell Biology Data Insights grant (2022-249319) to S.G. and P.P.. All funding bodies had no role in the design of the study and collection, analysis, and interpretation of data and in writing the manuscript.

**Conflict of Interest**:
The authors have no declared conflicts of interests.

**Author Contributions**:
Conceptualization: BG, WL, SCH
Formal Analysis: BG, WL
Funding acquisition: SG, KM, SCH
Project Administration: SG, KM, SCH
Supervision: SG, KM, SCH
Visualization: BG, WL, SHK
Writing – original draft: BG, WL, SCH
Writing – review & editing: BG, WL, SHK, PP, SG, KM, SCH

**Author ORCID**:
- Boyi Guo (https://orcid.org/0000-0003-2950-2349)
- Wodan Ling (https://orcid.org/0000-0001-7196-8543)
- Sang Ho Kwon (https://orcid.org/0000-0001-5328-0956)
- Pratibha Panwar (https://orcid.org/0000-0002-7437-7084)
- Shila Ghazanfar (https://orcid.org/0000-0001-7861-6997)
- Keri Martinowich (https://orcid.org/0000-0002-5237-0789)
- Stephanie C. Hicks (https://orcid.org/0000-0002-7858-0231)